\documentclass[apj]{emulateapj}
\usepackage{mathptmx}

\newcommand  \kms      {\ifmmode {\rm km\,s}^{-1} \else km\,s$^{-1}$\fi}
\newcommand  \cc       {\hbox{cm$^{-3}$}}
\newcommand  \cmii     {\hbox{cm$^{-2}$}}
\newcommand  \ergs     {\ifmmode {\rm erg\,s}^{-1} \else ergs s$^{-1}$\fi}
\newcommand  \ergcms   {\ifmmode {\rm erg\,cm}^{-2}\,{\rm s}^{-1}
                        \else erg\,cm$^{-2}$\,s$^{-1}$\fi}
\newcommand  \ergcmsA  {\ifmmode{\rm ergs\,cm}^{-2}\,{\rm s}^{-1}\,{\rm\AA}^{-1}
                        \else erg\,cm$^{-2}$\,s$^{-1}$\,\AA$^{-1}$\fi}
\newcommand  \ergcmsHz {\ifmmode{\rm ergs\,cm}^{-2}\,{\rm s}^{-1}\,{\rm Hz}^{-1}
                        \else erg\,cm$^{-2}$\,s$^{-1}$\,Hz$^{-1}$\fi}
\newcommand  \phcms    {\ifmmode {\rm photons\,cm}^{-2}\,{\rm s}^{-1}
                        \else photons\,cm$^{-2}$\,s$^{-1}$\fi}
\newcommand  \phcmsA   {\ifmmode {\rm photons\,cm}^{-2}\,{\rm s}^{-1}\,{\rm\AA}^{-1}
                        \else photons\,cm$^{-2}$\,s$^{-1}$\,\AA$^{-1}$\fi}

\newcommand  \mr       {MR~2251$-$178}

\journalinfo{The Astrophysical Journal, 6??: ??1--?14, 2004 August 10, astro-ph/0404565}
\slugcomment{Received 2003 October 2; accepted 2004 April 25}

\shorttitle{EVOLUTION OF THE IONIZED GAS IN \mr}

\shortauthors{KASPI ET AL.}

\begin{document}

\title{The Properties and the Evolution of the Highly Ionized Gas in \mr}
\author{
Shai~Kaspi,\altaffilmark{1} 
Hagai~Netzer,\altaffilmark{1}
Doron~Chelouche,\altaffilmark{1}
Ian~M.~George,\altaffilmark{2,3}
Kirpal~Nandra,\altaffilmark{4}
and
T.~J.~Turner\altaffilmark{2,3}
}

\altaffiltext{1}{School of Physics and Astronomy, Raymond and Beverly Sackler
Faculty of Exact Sciences, Tel-Aviv University, Tel-Aviv 69978, Israel; shai@wise.tau.ac.il}
\altaffiltext{2}{Laboratory for High Energy Astrophysics, NASA/Goddard Space
Flight Center, Code 662, Greenbelt, MD 20771.}
\altaffiltext{3}{Joint Center for Astrophysics, Physics Department, University
of Maryland, Baltimore County, 1000 Hilltop Circle, Baltimore, MD 21250.}
\altaffiltext{4}{Astrophysics Group, Imperial College London, Blackett 
Laboratory, Prince Consort Road, London SW7 2AZ, UK.}

\begin{abstract}
We present the first {\it XMM-Newton} observations of the radio-quiet
quasar \mr\ obtained in 2000 and 2002. The EPIC-pn spectra show a
power-law continuum with a slope of $\Gamma=1.6$ at high energies
absorbed by at least two warm absorbers (WAs) intrinsic to the source.
The underlying continuum in the earlier observation shows a ``soft
excess'' at low X-ray energies which can be modeled as an additional
power-law with $\Gamma=2.9$. The spectra also show a weak narrow
iron K$\alpha$ emission line. The high-resolution grating spectrum
obtained in 2002 shows emission lines from \ion{N}{6}, \ion{O}{7},
\ion{O}{8}, \ion{Ne}{9}, and \ion{Ne}{10}, as well as absorption
lines from the low-ionization ions of \ion{O}{3}, \ion{O}{4}, and
\ion{O}{5}, and other confirmed and suspected weaker absorption lines.
The \ion{O}{3}---\ion{O}{5} lines are consistent with the properties
of the emission line gas observed as extended optical [\ion{O}{3}]
emission in this source. The signal-to-noise of the 2000 grating
data is too low to detect any lines. We suggest a model for the
high-resolution spectrum which consist of two or three warm-absorber
(WA) components.  The two-components model has a high-ionization
WA with a column density of $10^{21.5}$--$10^{21.8}$ cm$^{-2}$
and a low-ionization absorber with a column density of $10^{20.3}$
cm$^{-2}$. In the three-components model we add a lower ionization
component that produce the observed iron M-shell absorption lines.
We investigate the spectral variations in \mr\ over a period of
8.5 years using data from {\it ASCA}, {\it BeppoSAX}, and {\it
XMM-Newton}. All X-ray observations can be fitted with the above two
power laws and the two absorbers. The observed luminosity variations
seems to correlate with variations in the soft X-ray continuum. The
8.5 year history of the source suggests a changing X-ray absorber
due to material that enters and disappears from the line-of-sight on
timescales of several months.  We also present, for the first time,
the entire {\it FUSE} spectrum of \mr . We detect emission from
\ion{N}{3}, \ion{C}{3}, and \ion{O}{6} and at least 4 absorption
systems in \ion{C}{3}, \ion{H}{1}, and \ion{O}{6}, one at $-$580 \kms\
and at least 3 others which are blended together and form a wide
trough covering the velocity range of 0 to $-$500 \kms . The general
characteristics of the UV and X-ray absorbers are consistent with an
origin in the same gas.
\end{abstract}


\section{Introduction}

\mr\ was the first quasar detected by X-ray observations (by {\it
Ariel~V\/} and {\it SAS-3}; Cooke et al. 1978, Ricker et al. 1978),
and also the first quasar where a warm absorber (WA) was suggested
to explain the X-ray spectrum based on {\it Einstein} observations
(Halpern 1984).  The X-ray flux of the source is variable on timescales
of $\sim 10$ days, e.g., the {\it EXOSAT} observations reported by
Pan et al. (1990). These authors found the column density of the
WA to vary and to correlate with the X-ray flux. Mineo \& Stewart
(1993) combined the earlier {\it EXOSAT} observations with a {\it
GINGA} observation from 1989 and argued that the spectrum could be
described by a power law with photon index $\Gamma\approx 1.7$ and
a WA model with a column density of $10^{22.2}$\,\cmii . Using this
model they found the ionization parameter to be strongly correlated
with the source luminosity.

A deep {\it ROSAT}/PSPC observation from 1993 was reported
by Komossa (2001) who modeled the 0.1--2.4~keV spectrum using
a WA with a column density of $10^{22.6}$\,\cmii .  \mr\ was
observed with {\it ASCA} ten times during 1993 and 1996.  These
observations seem to be consistent with a WA with a column density
of about $10^{21.3-21.7}$\,\cmii\ (e.g., Reynolds 1997, Otani et
al. 1998, Reeves \& Turner 2000, Morales \& Fabian 2002). \mr\
was also observed twice with {\it BeppoSAX} during 1998. The two
observations are separated by 5 months and show identical spectral
shape and flux. The WA column was found to be $10^{21.9}$~\cmii\
and the difference from previous observations was attributed to the
motion of the absorber across our line-of-sight (Orr et al. 2001).
In summary, the observed 2--10 keV flux of \mr\ covers the range of
$(1.7$--$5.1)\times10^{-11}$ \ergcms\ which translates to a 2--10
keV luminosity of $(1.7$--$5.2)\times10^{44}$~\ergcms\ (H$_{0}=70$
km\,s$^{-1}$\,Mpc$^{-1}$, $\Omega_{M}=0.3$, $\Omega_{\Lambda}=0.7$
and assuming a $\Gamma=1.6$ power law). The Galactic hydrogen column
density towards \mr\ has been derived from 21~cm measurements to be
$2.8\times10^{20}$~\cmii\ (Lockman \& Savage 1995).

\begin{deluxetable*}{ccccc}
\tablecolumns{5}
\tablewidth{288pt}
\tablecaption{Observation Log for \mr
\label{ascalog}}
\tablehead{
\colhead{Mission} &
\colhead{Sequence Number} &
\colhead{Date} &
\colhead{Time (ks)\tablenotemark{a}} &
\colhead{Rate (count\,s$^{-1}$)\tablenotemark{b}}
 }
\startdata
{\it ASCA } & 71035000 & 1993 Nov 11 & \phn  3.6 & $1.69\pm0.02$ \\
{\it ASCA } & 71035010 & 1993 Nov 16 & \phn  6.9 & $1.54\pm0.02$ \\
{\it ASCA } & 71035040 & 1993 Dec 12 & \phn  9.9 & $1.74\pm0.01$ \\
{\it ASCA } & 71035060 & 1993 Dec 14 & \phn  5.8 & $1.47\pm0.02$ \\
{\it ASCA } & 71035020 & 1993 Dec 19 & \phn  6.7 & $1.28\pm0.02$ \\
{\it ASCA } & 71035050 & 1993 Dec 24 & \phn  7.5 & $1.02\pm0.01$ \\
{\it ASCA } & 74028000 & 1996 May 26 &  17.8 & $0.65\pm0.01$ \\
{\it ASCA } & 74028010 & 1996 Jun 18 &  16.5 & $0.81\pm0.01$ \\
{\it ASCA } & 74028020 & 1996 Nov 27 &  15.9 & $0.57\pm0.01$ \\
{\it ASCA } & 74028030 & 1996 Dec 09 &  20.0 & $0.61\pm0.01$ \\
{\it BeppoSAX}  & 50556001 & 1998 Jun 14 & 47.5 &$0.419\pm0.002$\\
{\it BeppoSAX}  & 505560011& 1998 Nov 12 & 47.5 &$0.431\pm0.002$\\
{\it XMM-Newton}& 0112910301 & 2000 May 29 &\phn 3.5&$17.91\pm0.07$\phn\\
{\it XMM-Newton}& 0012940101 & 2002 May 18 & 44.7 &$\phn 7.11\pm0.01$  
\enddata
\tablenotetext{a}{Total effective exposure times determined from:
SIS0 selected data for {\it ASCA}, 
MECS data for {\it BeppoSAX},
and EPIC-pn data for {\it XMM-Newton}.}
\tablenotetext{b}{Count rates determined from:
SIS0 over the 0.5--10 keV band for {\it ASCA},
MECS over the 2--10 keV band for {\it BeppoSAX},
and EPIC-pn over the 0.2--11 keV band for {\it XMM-Newton}.}
\end{deluxetable*}

An Fe\,K$\alpha$ line was first suggested in a {\it
GINGA\/} observation of \mr, with an equivalent width (EW) of
$125^{+100}_{-105}$ eV (Mineo \& Stewart 1993).  This was later
confirmed by the {\it ASCA} and {\it BeppoSAX} observations with EW
of $190^{+140}_{-95}$ eV (Reynolds 1997; cf., Reeves \& Turner (2000)
found EW of $79\pm52$ eV) and $62^{+12}_{-25}$ eV (Orr et al. 2001),
respectively.

The redshift of \mr\ was determined by using nine optical narrow
emission lines (Bergeron et al. 1983) to be $z=0.06398\pm0.00006$
(we note that a few catalogs 
listed incorrect values, which has resulted in a 
variety of values quoted in the literature). The host galaxy
has a gaseous component with temperature of $\sim 3\times10^4$\,K
(derived from the [\ion{O}{3}] line ratio) and indications of low
abundances of Ne, O, and N (Bergeron et al. 1983). The galaxy is
surrounded by a giant \ion{H}{2} envelope which is observed via
[\ion{O}{3}]\,{$\lambda$}5007 emission. \mr\ was observed by {\it
HST} at three epochs (Monier et al. 2001 and references therein). The
spectrum shows clear Ly$\alpha$ and \ion{C}{4} absorption. Ganguly,
Charlton, \& Eracleous (2001) found the \ion{C}{4} doublet absorption
to vary with time, suggesting an intrinsic origin for this absorption.
The quasar is radio quiet with a radio flux of 16.2$\pm$0.7 mJy
(NVSS catalog -- Condon et al. 1998).
%
% R=5.2

This paper presents new {\it XMM-Newton} and {\it Far Ultraviolet
Spectroscopic Explorer (FUSE)} observations of \mr . We also carry out
an in-depth analysis of the 10 available {\it ASCA} observations and
the two {\it BeppoSAX} observations. We describe the data in \S~2,
perform an X-ray data temporal analysis in \S~3, and discuss the X-ray
spectral analysis in \S~4 and the UV spectral analysis in
\S~5. In \S~6 we elaborate on the implications of our results.

\section{Observations and Data Reduction}

This paper includes an extensive analysis of the historical UV
and X-ray spectra of \mr. The basic X-ray observations are listed in
Table~\ref{ascalog} and the data analysis is described in this section.

\subsection{\it XMM-Newton Observations}
\label{xmm_obs}

\mr\ was observed with {\it XMM-Newton} during 2002 May 18--19
for 64 ks. Data were reduced using the Science Analysis System
(SAS v5.3.0) in the standard processing chains as described in the
data analysis threads and the ABC Guide to {\it XMM-Newton} Data
Analysis.\footnote{http://heasarc.gsfc.nasa.gov/docs/xmm/abc} Source
data were extracted from circular regions of radius 30$\arcsec$ and
40$\arcsec$ for the EPIC-pn and EPIC-MOS, respectively. The EPIC-pn
was operated in the small window mode resulting with good exposure
time of 45 ks. The observed count rate ($\sim$7.3 counts\,s$^{-1}$
before background subtraction) was well below the pileup threshold
(130 counts\,s$^{-1}$, see {\it XMM-Newton} Users' Handbook).
For statistical purposes we binned the spectra to have at least
25 counts per bin. MOS CCDs were used in the large window mode
and the observed count rate ($\sim$2.0 counts\,s$^{-1}$ before
background subtraction) was just above the pileup threshold (1.8
counts\,s$^{-1}$). The MOS observations will not be discussed here.

\begin{figure*}
\centerline{\includegraphics[width=18.0cm]{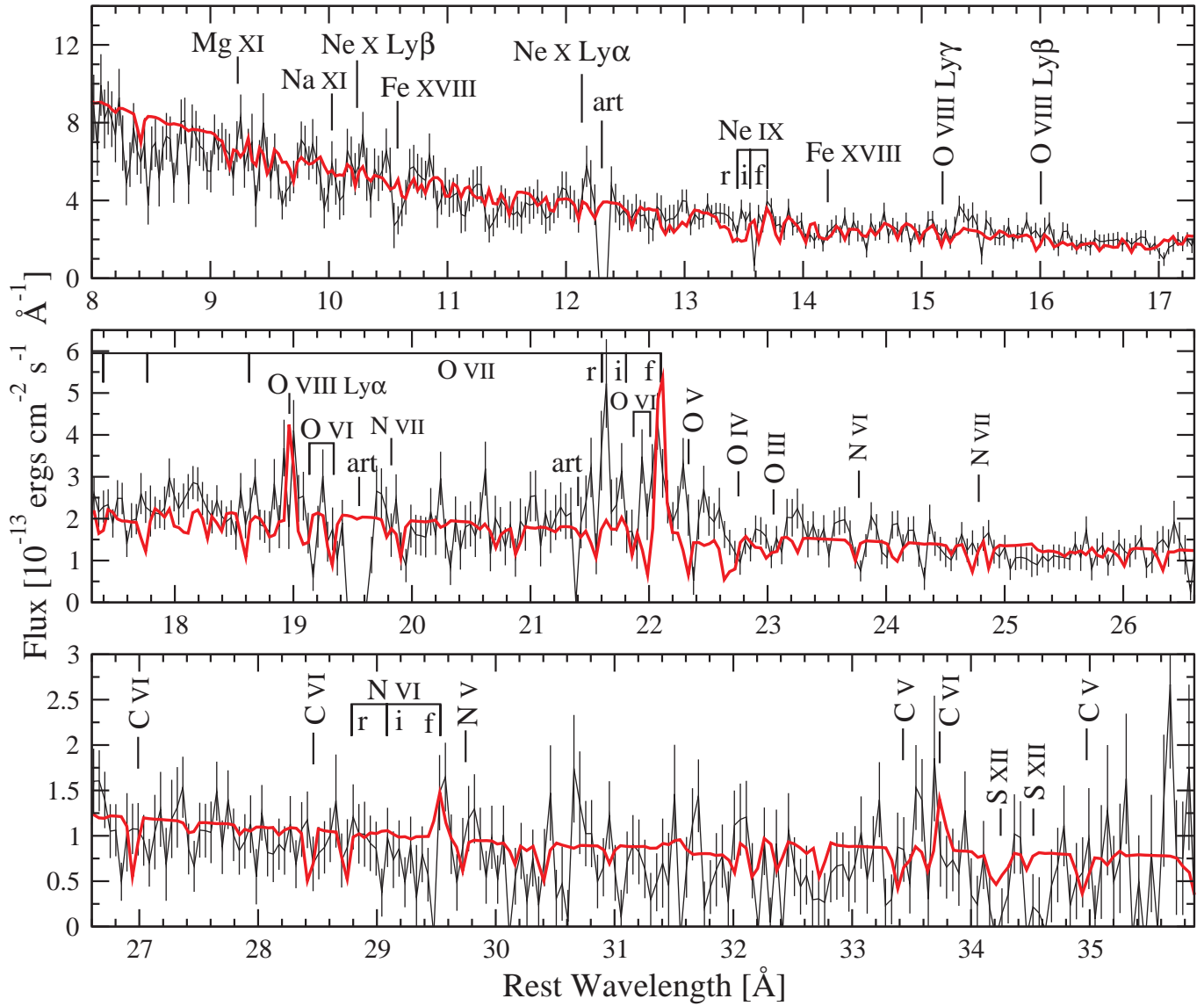}}
\caption{Combined RGS1 and RGS2 spectrum of \mr\ binned to $\sim 0.04$
 \AA . The spectrum has been corrected for Galactic
absorption and for the redshift of the source. The strongest
emission lines are due to the \ion{O}{7} triplet and \ion{O}{8}
Ly$\alpha$. Other suggested absorption and emission lines are marked.
Gaps in the spectrum due to chip gaps are marked as `art'. The 
three absorbers model discussed in \S~\ref{rgsfit} (red line) was convolved
with the RGS instrumental resolution and was also binned to 0.04~\AA .
\label{rgsspec} }
\vspace{-0.3cm}
\end{figure*}

The RGS1 and RGS2 were operated in the standard spectroscopy mode
resulting in a good exposure time of 63 ks for each.  Background
extraction is performed with the SAS using regions adjacent to those
containing the source in the spatial and spectral domains. The spectra
were extracted into bins of $\sim$\,0.04 \AA\ in width (4 times the
default bin width) in order to increase the signal-to-noise ratio.
To flux calibrate the RGS spectra we divided the counts by the
exposure time and by the effective area at each wavelength.
Each flux-calibrated spectrum was also corrected for Galactic
absorption and the two spectra combined into an error-weighted
mean. At wavelengths where the RGS2 bins did not match exactly the
wavelength of the RGS1 bins, we interpolated the RGS2 data to enable
the averaging. This final spectrum is shown in Figure~\ref{rgsspec}.

\mr\ was also observed with {\it XMM-Newton} during the validation and
verification phase of the telescope during 2000 May 29. We retrieved
the data from the {\it XMM-Newton} archive and reduced it in the same
way as described above for the 2002 observation.  Unfortunately, due
to the operating modes used, neither of the EPIC-MOS detectors contain
any useful data from \mr\ during this observation.  The EPIC-pn was
operated in the small window mode resulting in good exposure time of
3.5 ks. The observed count rate ($\sim$18.1 counts\,s$^{-1}$ before
background subtraction) was well below the pileup threshold. The RGS1
and RGS2 were operated in the standard spectroscopy mode during two
distinct exposures, each of about 6 ks (which is the same as the good
exposure time accumulated).

\begin{figure*}
\centerline{\includegraphics[width=18.0cm]{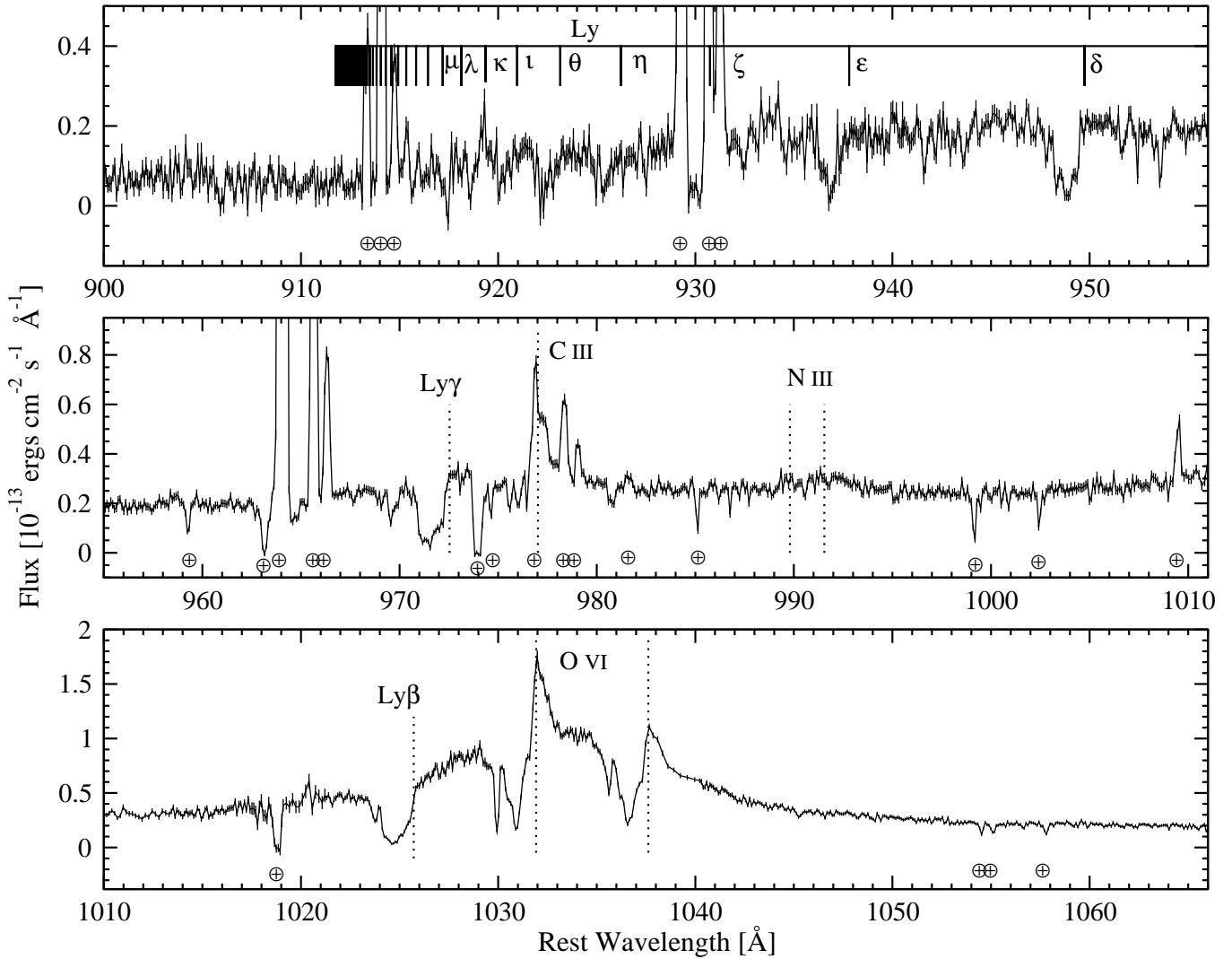}}
\caption{{\it FUSE} spectrum of \mr\ binned to $\sim 0.1$ \
\AA . Identified intrinsic emission lines are marked at their
theoretical position above the spectrum and identified airglow
and Galactic lines are marked below the spectrum with~$\oplus$. The
intrinsic \ion{H}{1} absorptions lines are marked at their theoretical
expected position up to the ionization edge.
\label{fusespec} }
\vspace{0.4cm}
\end{figure*}

\subsection{\it ASCA Observations}

As listed in  Table~\ref{ascalog}, the {\it ASCA} archive contains 10
observations of \mr.  The data were screened in the same manner as
used for the {\em Tartarus Database} (e.g., Turner et al. 1998). We
briefly describe this procedure here. ASCASCREEN/XSELECT (v0.45)
was used for screening together with the criteria\footnote{With the
exception of the pixel threshold for each SIS that was automatically
calculated on a file-by-file basis within ASCASCREEN.} given in Nandra
et al. (1997). In the case of the {\it ASCA\/} SIS data, 'hot' and
'flickering' pixels were removed using the standard algorithm and only
SIS grades 0, 2, 3, and 4 events were included in the analysis. The
original pulse-height assignment for each event was converted to
a pulse-invariant (PI) scale using SISPI (v1.1). In the GIS data,
hard particle flares were rejected using the so-called 'HO2' count
rate, and standard rise-time rejection criteria were employed. The
effective exposure times resulting from these criteria are listed
in Table~\ref{ascalog}.

The spectral analysis for each of the 10 {\it ASCA}
observations was performed on the data from all four instruments
simultaneously. Different relative normalizations were allowed to
account for (small) uncertainties in the determination of the effective
area of each instrument. We also corrected the {\it ASCA} 1996 data for
the SIS degradation as indicated by Yaqoob et al. (2000). Individual
spectra were binned in energy to contain a minimum of 20 counts per
bin, and hence allowing meaningful $\chi^2$ minimization.  Fits to the
data were carried out with XSPEC using our own models, as described
below. Quoted uncertainties on the parameters refer to the 90\%
confidence level.

\subsection{\it BeppoSAX Observations}

{\it BeppoSAX} observed \mr\ at two epochs: 1998 June 14--18 and 1998
November 12--16. The observations and their analysis are described
in Orr et al. (2001). In the following analysis we use the data and
the calibrations supplied by the HEASARC archive. We used only the
data obtained with the Low-Energy Concentrator Spectrometer (LECS;
0.1--4 keV) and the Medium-Energy Concentrator Spectrometer (MECS;
1.8--10 keV).  Again we use XSPEC, and allow different relative
normalizations to account for uncertainties in the effective areas
of the instruments.

\subsection{\it FUSE Observations}

\mr\ was observed with {\it FUSE} during 2001 June 20--21.
The observation was carried out using the LWRS aperture and is
$\sim 50$ ks in duration.  Only a small part of the spectrum around
the \ion{O}{6} absorption has been published to date (Wakker et
al. 2003). Thus we have extracted the raw data from the {\it FUSE}
archive, and reduced it using the {\it FUSE} software (CalFUSE v2.2.2
and FUSE IDL tools version of 2002 July). The {\it FUSE} spectrum is
shown in Figure~\ref{fusespec}.

\section{Temporal Analysis of the X-ray Spectra}

\subsection{\it ASCA}

Light curves were constructed for the source and background regions for
all {\it ASCA} observations, in several different energy ranges. To
increase the signal-to-noise ratio, the light curves from each pair
of SIS and GIS detectors were combined. The light curves were then
rebinned on a variety of timescales.

\begin{figure}
\centerline{\includegraphics[width=8.5cm]{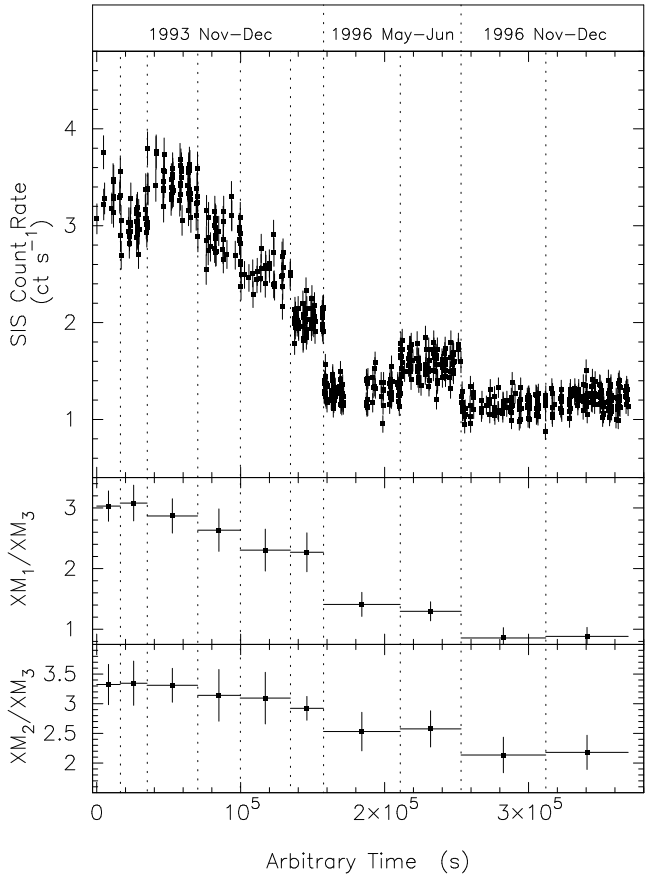}}
\caption{{\it Upper panel}: summed SIS0 and SIS1 light curve over
the 0.5--10 keV band for the {\it ASCA} observations of \mr\ using a
128~s bin size. {\it Lower panels}: the mean `softness ratio' of each
observation using the 0.5--1.2~keV (XM$_1$), 1.5--3.5~keV (XM$_2$),
and 4.0--10.0~keV (XM$_3$)bands. Note that the spectrum is softer
when the source is brighter.
\label{sislc} }
\vspace{0.1cm}
\end{figure}

The combined SIS light curves in the 0.5--10 keV band are shown
in Figure~\ref{sislc}. There is no evidence for short timescale
variability within any of the observations. This is confirmed by a lack
of significant `excess variance' (Turner et al. 1999 and references
therein) in any of the observations, with upper limits of typically
$\sigma_{\rm rms}^{2}\approx 3\times 10^{-3}$ (at 90\% confidence).
Such a lack of a short timescale variability is consistent with the
anti-correlation between the excess variance and the luminosity
found by Nandra et al. (1997). Clear variability is seen between
observations, with a maximum flux change of a factor $\sim$\,3 over
3 years. The smallest amplitude variation on the shorter timescale is
a 20\% flux decrease in 5 days between the fifth and the sixth {\it
ASCA} observations.

We have also constructed light curves in the 0.5--1.2 keV (XM$_1$),
1.5--3.5 keV (XM$_2$), and 4.0--10.0 (XM$_3$) bands of Netzer, Turner,
\& George (1994). In the lower panels of Figure~\ref{sislc} we plot the
mean XM$_1$/XM$_3$ and XM$_2$/XM$_3$ ratios for each observation. We
find no statistically significant variation in softness ratio during
each of the three epochs (the apparent decrease in the XM$_1$/XM$_3$
ratio during the 1993 Nov--Dec epoch is significant at only the
75\% confidence level). However, both ratios exhibit statistically
significant variability (98\% confidence) {\it between the epochs} in
a manner suggesting that the spectrum becomes softer when the source
luminosity increases. These variations are discussed in \S\ref{history}
and \S\ref{historydiss}.

\subsection{\it XMM-Newton and BeppoSAX}

We examined the EPIC-pn and the two EPIC-MOS background-subtracted
light curves of \mr . No significant flux variation is detected during
the 64 ks observation. This is consistent with the {\it ASCA}
observations which show no variations on timescales of less than a day.

The EPIC-pn count rate from the 2000 observation ($17.9\pm0.07$
counts\,s$^{-1}$) is 2.5 times higher than the EPIC-pn data from
2002 ($7.1\pm0.01$ counts\,s$^{-1}$). Table~\ref{pnsoftrat} gives the
softness ratios, as defined above, for the two EPIC-pn observations. We
find clear variations in both softness ratios.

Orr et al. (2001) examined the light curves for the 1998 June
and November {\it BeppoSAX} observations as well as the hardness
ratios. They find the count rates and hardness ratios to be similar
at the two epochs. While the June light curve is well fitted with a
constant count rate Orr et al. (2001) find the November observation to
better fit with a slowly decreasing linear function (minus $\sim 15$\%
in 70 hours) although a constant count rate cannot be ruled out.

\begin{deluxetable}{ccc}
\tablecolumns{3}
\tablewidth{170pt}
\tablecaption{Softness Ratios for {\it XMM-Newton} Observations
\label{pnsoftrat}}
\tablehead{
\colhead{Date} &
\colhead{XM$_1$/XM$_3$} &
\colhead{XM$_2$/XM$_3$}
 }
\startdata
2000 May 29 & $3.964\pm0.060$ & $2.352\pm0.038$ \\
2002 May 18 & $2.549\pm0.016$ & $2.014\pm0.013$   
\enddata
\end{deluxetable}

\begin{figure}
\centerline{\includegraphics[width=8.5cm]{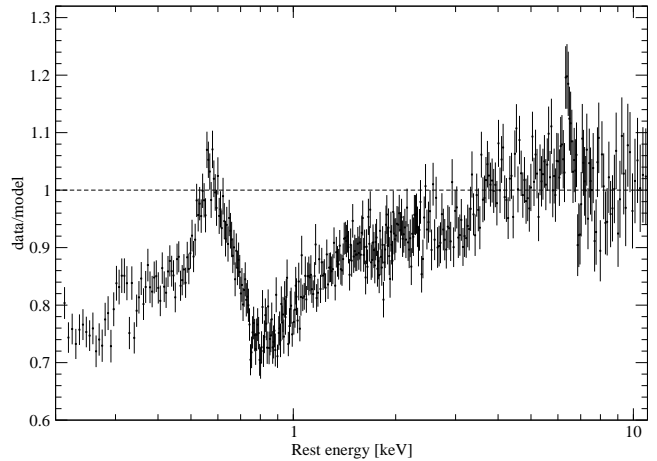}}
\caption{Ratio of the 2002 {\it XMM-Newton} EPIC-pn data to a power
law model (including Galactic absorption) fitted to the 3--11 keV
band. Excess absorptions is evident near the \ion{O}{7} and the
\ion{O}{8} absorption edges and at $E <0.5$\,keV.
\label{datamo2002} }
\vspace{-0.1cm}
\end{figure}

\section{Spectral Analysis of the X-ray Data}

We have carried out an extensive spectral analysis of the low
resolution {\it ASCA, BeppoSAX} and {\it XMM-Newton} spectra of \mr\
as well as the high resolution RGS spectra. We first consider the
high signal-to-noise, broad-band and RGS spectra obtained using {\it
XMM-Newton} in 2002, and describe a multi-component model that
is consistent with the data (\S\ref{2002-broad} and \ref{rgsfit}).
This model is then compared to the {\it XMM-Newton} data obtained
during 2000 (\S\ref{softexcess} and \ref{2000-rgs}) and to the
historical X-ray datasets in \S\ref{history}.

\subsection{Analysis of the 2002 Broad Band X-ray Spectrum}
\label{2002-broad}

We first fitted the 0.2--11 keV EPIC-pn spectrum of \mr\ obtained
in 2002 with a single Galactic absorbed power law. This gives
a poor description of the data ($\chi^{2}_{\nu}\approx 1.9$)
in agreement with previous analysis of the source. The main
discrepancy is at low energies, hence we repeated the fit
for only the 3--11 keV band (excluding the 4.5--7.5 keV
rest-frame band where contamination from the Fe\,K$\alpha$
line might be present).  We find $\Gamma=1.535\pm0.024$
and normalization of $(4.00^{+0.10}_{-0.14})\times10^{-3}$
photons\,cm$^{-2}$\,s$^{-1}$\,keV$^{-1}$ with $\chi^{2}_{\nu}=1.02$.
We show an extrapolation of this continuum to lower energies in
Figure~\ref{datamo2002}. The presence of an excess absorption around
the \ion{O}{7} and \ion{O}{8} edges, at 0.7--0.9 keV, is clear
and indicates a WA component. Excess absorption is also evident
at energies below 0.5 keV, which is indicative of an additional,
less-ionized absorber.

\begin{deluxetable*}{cccccccc}
\tablecolumns{8}
\tablewidth{420pt}
\tablecaption{Spectral Fits to Low Resolution X-ray Observations
\label{xrayfit}}
\tablehead{
\colhead{Mission name} &
\colhead{Date} &
\colhead{$\log$ Column\tablenotemark{a}} &
\colhead{$\log U_{\rm OX}$\tablenotemark{b}} &
\colhead{Norm ($\Gamma=2.9)$\tablenotemark{c}} &
\colhead{Norm ($\Gamma=1.6)$\tablenotemark{d}} &
\colhead{Flux\tablenotemark{e}} &
\colhead{$\chi_{\nu}^{2}$} \\
\colhead{(1)} &
\colhead{(2)} &
\colhead{(3)} &
\colhead{(4)} &
\colhead{(5)} &
\colhead{(6)} &
\colhead{(7)} &
\colhead{(8)}  \\
\hline 
\multicolumn{8}{c}{Fits to the 0.1--10.0 keV band}
}
\startdata
{\it BeppoSAX}  & 1998 Jun 14 & 21.8 & $-1.86^{+0.09}_{-0.12}$ &  $2.34^{+0.51}_{-0.40}$    & $6.46^{+0.24}_{-0.23}$  &  2.04 & 1.25 \\
{\it BeppoSAX}  & 1998 Nov 12 & 21.8 & $-1.98^{+0.13}_{-0.19}$ &  $2.45^{+0.57}_{-0.53}$    & $7.03^{+0.27}_{-0.28}$  &  2.22 & 1.24 \\
{\it XMM-Newton}\tablenotemark{f}& 2000 May 29 & 21.5 & $ -1.76^{+0.07}_{-0.05}$ &  $2.45^{+0.24}_{-0.21}$    & $7.11^{+0.18}_{-0.20}$  &  2.27 & 1.15 \\
{\it XMM-Newton}\tablenotemark{f}& 2002 May 18 & 21.5 & $ -1.79^{+0.03}_{-0.03}$ &  $0.00^{+0.01}_{-0.00}$    & $4.17^{+0.03}_{-0.03}$  &  1.28 & 1.21 \\ 
\cutinhead{Fits to the 0.5--10.0 keV band}
{\it ASCA}      & 1993 Nov 11 & 21.5 & $ -1.80^{+0.80}_{-0.46}$ &  $4.02^{+1.50}_{-1.34}$    & $10.41^{+0.35}_{-0.46}$  &  3.39 & 0.96 \\
{\it ASCA}      & 1993 Nov 16 & 21.5 & $ -2.02^{+0.76}_{-0.11}$ &  $5.73^{+0.52}_{-1.61}$    & $ 8.11^{+1.01}_{-0.15}$  &  2.72 & 0.99 \\
{\it ASCA}      & 1993 Dec 12 & 21.5 & $ -2.22^{+0.21}_{-0.20}$ &  $6.17^{+0.74}_{-1.24}$    & $ 9.49^{+0.45}_{-0.34}$  &  3.17 & 0.98 \\
{\it ASCA}      & 1993 Dec 14 & 21.5 & $ -2.12^{+0.30}_{-0.58}$ &  $3.95^{+0.56}_{-1.28}$    & $ 8.37^{+0.44}_{-0.28}$  &  2.85 & 1.02 \\
{\it ASCA}      & 1993 Dec 19 & 21.5 & $ -2.33^{+0.26}_{-0.27}$ &  $2.60^{+0.74}_{-0.83}$    & $ 8.00^{+0.46}_{-0.41}$  &  2.65 & 0.92 \\
{\it ASCA}      & 1993 Dec 24 & 21.5 & $ -1.75^{+0.47}_{-0.73}$ &  $0.87^{+1.77}_{-0.74}$    & $ 6.62^{+0.30}_{-0.44}$  &  2.20 & 0.96 \\
{\it ASCA}      & 1996 May 26 & 21.8 & $ -2.12^{+0.09}_{-0.32}$ &  $0.00^{+0.06}_{-0.00}$    & $ 5.45^{+0.20}_{-0.21}$  &  1.74 & 0.96 \\
{\it ASCA}      & 1996 Jun 18 & 21.8 & $ -2.27^{+0.13}_{-0.12}$ &  $0.00^{+0.03}_{-0.00}$    & $ 6.75^{+0.19}_{-0.10}$  &  2.14 & 1.03 \\
{\it ASCA}      & 1996 Nov 27 & 22.1 & $ -2.27^{+0.10}_{-0.11}$ &  $0.00^{+0.06}_{-0.00}$    & $ 5.63^{+0.18}_{-0.10}$  &  1.82 & 1.01 \\
{\it ASCA}      & 1996 Dec 09 & 22.1 & $ -2.19^{+0.07}_{-0.14}$ &  $0.09^{+0.40}_{-0.09}$    & $ 6.08^{+0.34}_{-0.15}$  &  1.95 & 1.00 \\
{\it BeppoSAX}  & 1998 Jun 14 & 21.8 & $ -1.86^{+0.17}_{-0.20}$ &  $2.45^{+1.03}_{-1.10}$    & $ 6.42^{+0.37}_{-0.37}$  &  2.03 & 1.26 \\
{\it BeppoSAX}  & 1998 Nov 12 & 21.8 & $ -1.88^{+0.20}_{-0.21}$ &  $1.65^{+1.10}_{-1.21}$    & $ 7.25^{+0.42}_{-0.43}$  &  2.27 & 1.23 \\
{\it XMM-Newton}& 2000 May 29 & 21.5 & $ -1.80^{+0.07}_{-0.05}$ &  $2.62^{+0.67}_{-0.60}$    & $7.08^{+0.28}_{-0.35}$  & 2.27 & 1.12 \\
{\it XMM-Newton}& 2002 May 18 & 21.5 & $ -1.83^{+0.03}_{-0.03}$ &  $0.00^{+0.04}_{-0.00}$    & $4.19^{+0.03}_{-0.03}$  & 1.28 & \phn\phn\phn1.16 
\enddata
\tablecomments{Fits use two power laws with fixed slopes of $\Gamma=1.6$
(normalization given in column 6) and $\Gamma=2.9$ (normalization given
in column 5), a photoionized WA with ionization parameter as given in
column (4) and column density fixed to the value given in column (3),
Galactic neutral absorber fixed at $2.8\times10^{20}$\,\cmii\
and intrinsic neutral absorber with column density of
$2\times10^{20}$\,\cmii . All fits are over the noted band,
excluding the rest frame 5.0--7.5 keV band.  Quoted uncertainties on
the parameters are at the 90\% confidence level. Several of these
fits are presented in Figure~\ref{softxmodels} and \ref{xmodels}.
The fits are discussed in \S~\ref{history} which also explain how the column densities
for the observations were determined.}
\tablenotetext{a}{Log of the column density in units of \cmii .}
\tablenotetext{b}{Ionization parameter, $U_{\rm OX}$, defined over
the 0.538--10 keV range.}
\tablenotetext{c}{Normalization at 1 keV in units of
$10^{-3}$\,photons\,cm$^{-2}$\,s$^{-1}$\,keV$^{-1}$ for the
power law with photon index fixed to $\Gamma=2.9$.}
\tablenotetext{d}{Normalization at 1 keV in units of
$10^{-3}$\,photons\,cm$^{-2}$\,s$^{-1}$\,keV$^{-1}$ for the
power law with photon index fixed to $\Gamma=1.6$.}
\tablenotetext{e}{Observed 4--10 keV flux in units of
$10^{-11}$\,\ergcms .}
\tablenotetext{f}{Fits for these {\it XMM-Newton} observations are
over the 0.2--11 keV band.}
\end{deluxetable*}

We used ION2003, the 2003 version of the ION photoionization code
(Netzer 1996; Netzer et al. 2003) in order to model the WA. For
simplicity, in all calculations we assume constant density gas with
a density of $10^5$ \cc , which is low enough to avoid complications
due to collisional de-excitation and line transfer yet large enough
to assume ``thin-shell'' geometry.  The relevant parameters of the
model are $U_{\rm OX}$ (the oxygen ionization parameter defined
over the range of 0.538--10 keV), the column density $N_{\rm H}$
(in units of \cmii), the gas composition (assumed to be solar and
specified in Netzer et al. 2003 Table 2), and the covering fraction.

We used ION2003 to fit the 2002 EPIC-pn data assuming power law, with
a slope fixed to the value found earlier ($\Gamma=1.54$), attenuated by
the Galactic column density, and two generic absorption components: one
with a ``typical'' WA properties and one which is much less ionized.
Fitting the spectrum with this model yields some excess emission around
0.5 keV regardless of the exact values of $U_{\rm OX}$ and $N_{\rm H}$.
We interpret this excess as due to emission of the \ion{O}{7} triplet
and the \ion{O}{8} Ly$\alpha$ lines. Hence, we added to the model an
emission component constrained to have the WA ionization parameter and
column density.  Having all these components, we obtained the following
solution: For the WA component we find log($U_{\rm OX})=-1.78\pm0.05$,
$N_{\rm H}=10^{21.51\pm0.03}$ \cmii\ and a line-of-sight covering
factor of 0.8. For the emission we find a global covering factor of
0.3. For the less ionized absorber we find that it can be fitted by a
neutral absorber (in addition to the Galactic one) with $N_H\approx
10^{20.3}$ \cmii . Since \mr\ is at low redshift this column can be
interpret either as an additional galactic absorber or as neutral gas
intrinsic to the source.  The low ionization absorber can also be
modeled as a combination of low ionization absorber with log($U_{\rm
OX})\approx-4$ and $N_{\rm H}\approx10^{20.3\pm0.03}$ \cmii\ and a
neutral absorber of $N_H\approx 10^{20.06}$ \cmii (we show below
that such low ionization component is required by the RGS data).
Both cases give equally good fits and the statistical analysis for
all absorbers and emitters yield $\chi^{2}_{\nu}\approx 1.14$ for both.

\subsubsection{Fe\,K$\alpha$ Line}

The 2002 {\it XMM-Newton} EPIC-pn data suggest a very weak
Fe\,K$\alpha$ line. Using the continuum from Table~\ref{xrayfit},
and fixing the line energy to 6.4 keV, we find a narrow line
with a width of $\sigma=0.099^{+0.052}_{-0.090}$ keV and a flux of
$(1.38\pm0.51)\times10^{-5}$ \phcms . The EW of the line is $53\pm20$
eV, consistent with previous studies of this source (see \S\,1).
We found no indication for a broad component to the Fe\,K$\alpha$
line. This line will not be discussed any further.

We note on a narrow absorption feature at around rest frame energy of
$\sim 7$ keV (see Figure~\ref{datamo2002}). We fitted this feature with
a Gaussian with a width fixed to the instrumental resolution and find
its rest frame energy to be $6.97\pm0.11$ keV and a normalization of
$(-6.4^{+4.4}_{-1.3})\times10^{-6}$ \phcms. The EW of this Gaussian
is $-28^{+20}_{-5}$ eV. The energy of this feature is consistent with
the \ion{Fe}{26} Ly$\alpha$ line.

\subsection{Analysis of the 2002 RGS Spectrum}
\label{rgsfit}

The 2002-RGS spectrum of \mr\ shows several absorption and emission
lines with wavelengths that are consistent with the systemic velocity,
given the RGS resolution (0.04~\AA, corresponding to 1000 \kms\ at
12~\AA\ and 400 \kms\ at 30~\AA ).  Despite the low signal-to-noise
ratio (S/N; of order 3--4 at around 20~\AA ) evidence can be
seen for emission lines from \ion{N}{6}, \ion{O}{7}, \ion{O}{8},
\ion{Ne}{9}, and \ion{Ne}{10}. Absorption lines are seen from the
low ionization ions of \ion{O}{3}, \ion{O}{4}, and \ion{O}{5} as well
as many higher ionization species. The strongest of these lines are
marked in Figure~\ref{rgsspec} together with many other lines whose
detection is less certain due to the low S/N.  The wavelengths of the
\ion{O}{3}--\ion{O}{6} lines are those used by Netzer et al. (2003). A
full line list is given in Table~\ref{linetable} where we differentiate
between lines that are identified with high certainty and these that
we regard as possible identification, due to the poor S/N.  We also
detect bound-free absorption due to the \ion{O}{7} and \ion{O}{8}
edges and a noticeable curvature of the spectrum over the wavelength
band of 15--17\AA.

The RGS spectrum also shows several features which we suspect to be
artifacts. The strong emission-like feature at $\sim 35.7$ \AA\ is
probably an artifact caused by high background level and the proximity
to the edge of the CCD. Several absorption-like features, at around
10--11 \AA , are similar in shape and intensity to other absorption
features and we suspect that some of those are due to \ion{Fe}{17}--\ion{Fe}{19}.
However, in this part of the spectrum there is only one CCD (RGS2)
and we cannot confirm their reality by comparing the two RGS spectra.

Modeling of the 2002 RGS spectrum was done in two steps.  First we
experimented with a two component absorber, similar to the one
discussed in \S~\ref{2002-broad}. This involves a highly-ionized
absorber and a second absorber of much lower-ionization. The
highly-ionized WA has a large column density and is responsible for
the bound-free absorption edges and the \ion{O}{7} and \ion{O}{8}
emission lines. The less ionized component has a lower column
density and is responsible for the \ion{O}{3} and \ion{O}{4}
absorption lines.  The absorption lines in both components are probably
narrower than the instrumental resolution and we assumed that they
can be characterized by a turbulent velocity of $b\approx 200$ \kms.
The modeling assumes that each of the components can be represented
by a single cloud (``shell''). Thus, the gas on the line of sight
produces the absorption features and the gas outside the line of
sight produces the emission lines.

\begin{deluxetable}{ccc}
\tablecolumns{3}
\tablewidth{172pt}
\tablecaption{Identified and suspected lines in the 2002 RGS spectrum
\label{linetable}}
\tablehead{
\colhead{Ion} &
\colhead{Rest wavelength [\AA ]} &
\colhead{Confidence\tablenotemark{a}}
 }
\startdata
\ion{C}{6}    &   33.737  &  2   \\  
\ion{N}{5}    &   29.747  &  2   \\  
\ion{N}{6}    &   29.541  &  1    \\  
\ion{N}{6}    &   28.787  &  2   \\      
\ion{C}{6}    &   28.465  &  2   \\  
\ion{C}{6}    &   26.990  &  2   \\  
\ion{N}{7}    &   24.782  &  2   \\  
\ion{N}{6}    &   23.771  &  1    \\  
\ion{O}{3}    &   23.051  &  1   \\  
\ion{O}{4}    &   22.729  &  1    \\  
\ion{O}{5}    &   22.334  &  1   \\     
\ion{O}{7}    &   22.101  &  1   \\  
\ion{O}{6}    &   22.007  &  1   \\  
\ion{O}{6}    &   21.788  &  1   \\  
\ion{O}{7}    &   21.602  &  1   \\  
\ion{N}{7}    &   19.826  &  2   \\  
\ion{O}{6}\tablenotemark{b}    &   19.341  &  2   \\  
\ion{O}{6}    &   19.135  &  1   \\  
\ion{O}{8}    &   18.966  &  1   \\  
\ion{O}{7}    &   18.627  &  2   \\ 
\ion{O}{8}    &   16.006  &  2   \\  
\ion{Fe}{18}   &  14.208  &  1  \\
\ion{Ne}{9}    &  13.700  &  1   \\  
\ion{Ne}{9}    &  13.447  &  1   \\     
\ion{Ne}{10}   &  12.134  &  1   \\  
\ion{Fe}{18}   &  10.579  &  1  \\
\ion{Ne}{10}   &  10.238  &  2   \\  
\ion{Na}{11}   &  10.025  &  1   \\ 
\ion{Fe}{19}   &   9.695  &  1  \\
\ion{Mg}{11}   &   9.231  &  2   \\  
\enddata
\tablenotetext{a}{Confidence mark for the identified lines: 1 --
line identified with high confidence; 2 -- line identified with
low confidence.}
\tablenotetext{b}{Blended with \ion{N}{7} at 19.361 \AA .}
\end{deluxetable}

We first used the model parameters derived form the EPIC-pn data and
calculated theoretical spectra for this gas.  We note that according
to Kirsch (2003) the EPIC and RGS agree within $\pm 20 $\% in the
normalization, and individual fitting shows a significant steeper
slope for the EPIC (see also den Herder et al. 2003 and Blustin et
al. 2002). This effect seems to be present in our data and hence we
do not require complete agreement between the slopes derived from
fitting the RGS and the PN spectra. Given those uncertainties, we
found a good fit for the RGS continuum with our two-component model
for a power law of photon index $\Gamma=1.5$.

We experimented with a range of parameters around the values found for
the EPIC-pn. The parameters we found to fit best for the two absorbers
in the RGS spectrum are: log($U_{\rm OX})=-1.68$ and $N_{\rm H}$ in
the range of $10^{21.5}$--$10^{21.8}$ \cmii\ for the high-ionization
WA component and log($U_{\rm OX})=-4.0$ and $N=10^{20.3}$ \cmii\
for the low-ionization component.  We also assume, based on the UV
measurement (see \S\ref{hiresfuse}) and in agreement with the EPIC-pn
model, a line of sight covering factor of 0.8 and an undetermined
outflow velocity which is taken to be 300 \kms .  We note that the
less-ionized gas does not contribute anything to the observed X-ray
emission lines and the required global ($4 \pi$) covering factor for
the emitting gas is $\sim 0.4$.

The two-component model fits the general continuum shape, the
\ion{Ne}{9} triplet, the \ion{O}{8} Ly$\alpha$, the \ion{O}{7}
forbidden line, and the absorption of \ion{O}{4} and \ion{O}{3}.
A major discrepancy is the underestimation of the \ion{O}{7} resonance
emission line at 21.6 \AA . A similar phenomenon has been observed
in NGC\,3783 where the \ion{O}{7} line is underpredicted in the
best-fitting model of Netzer et al (2003). A possible explanation may
be a complex optical depth structure for this optically thick line. For
example, the lateral optical depth (which cannot be observed and is
a function of the geometry) may be smaller than the line-of-sight
optical depth used in the calculations. As a results, line photons
can escape more easily in some directions increasing, in this way,
the emission line intensity. Such a situation may arise in conical
type flows where the cone lateral dimension is  smaller than its
height. The two-component fit is severely limited by the poor S/N
of the grating observations. Nevertheless, it shows that the ionized
(line-of-sight) absorber and the ionized emitter are consistent with
being the same gas.

\begin{figure}
\centerline{\includegraphics[width=8.5cm]{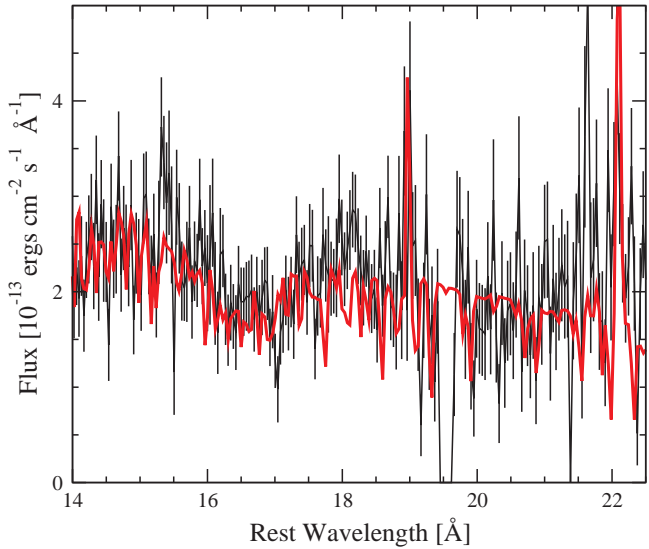}}
\caption{The RGS spectrum overplotted with the three absorbers model
discussed in  \S~\ref{rgsfit} (red line).  The three absorbers model
produces a noticeable UTA feature around 16--17 \AA\ and thus a better
fit compared with the two absorber model. However, it fails to explain
the EPIC-pn spectrum.
\label{threecompmo} }
\end{figure}

The next step includes a three component absorber. The main motivation
for this is the fact that the RGS data around 16--17~\AA\ clearly falls
below the two-component model. The excess absorption is probably caused
by the unresolved transition array (UTA) of iron M-shell lines (Behar,
Sako, \& Kahn 2001) which has been observed in several other AGNs
(see, e.g., Netzer et al. 2003 for the case of NGC\,3783 and Netzer
2004 for a general discussion). Our photoionization code includes all
these lines but the two-component WA produces too shallow a feature
at too short a wavelength. We find that an additional shell with a
column density of $10^{21.3}$ \cmii\ and log$(U_{\rm OX})=-2.6$ can
significantly improve the fit. This component produces a noticeable UTA
feature and contributes also to the observed \ion{O}{7} emission. This
requires lowering the emission from the high-ionization component
by about 20\% to produce an adequate fit to all emission lines.
Adding this component force us to increase the ionization parameter
of the highly ionized gas (the one with column density of $10^{21.8}$
\cmii ) to log$(U_{\rm OX})=-1.4$. We note that the mean $U_{\rm OX}$
of these two WA components is the same as the one found earlier in
the two-component model.
The three-component model is compared with the RGS data in
Figure~\ref{rgsspec} on a wavelength scale where all the features
can be seen. In Figure~\ref{threecompmo} we show a comparison on
a reduced wavelength scale to emphasize the UTA range. The UTA fit
utilizes the improved dielectronic recombination rates of Netzer (2004)
and is in good agreement with the 15--17 \AA\ spectrum.

There are two problems with the three-component model related to its
agreement with the EPIC-pn data. First, we could not find a model
which explains the UTA feature and is also consistent with the EPIC-pn
spectrum. In particular, while we are convinced in the presence of
a UTA feature, there is no way to assess the exact column density of
the relevant ions (\ion{Fe}{7}--\ion{Fe}{12} and \ion{O}{7}), in the
intermediate $U_{\rm OX}$ component, given the S/N of the present data.
Second, the chosen column of $10^{21.8}$ \cmii\ for the high $U_{\rm
OX}$ component is about the maximum which is still consistent with
the EPIC-pn observation. Yet, some strong features in the RGS spectrum
seem to require even a larger absorbing column density.

\subsection{Analysis of the 2000 Broad Band X-ray Spectrum}
\label{softexcess}

\begin{figure}
\centerline{\includegraphics[width=8.5cm]{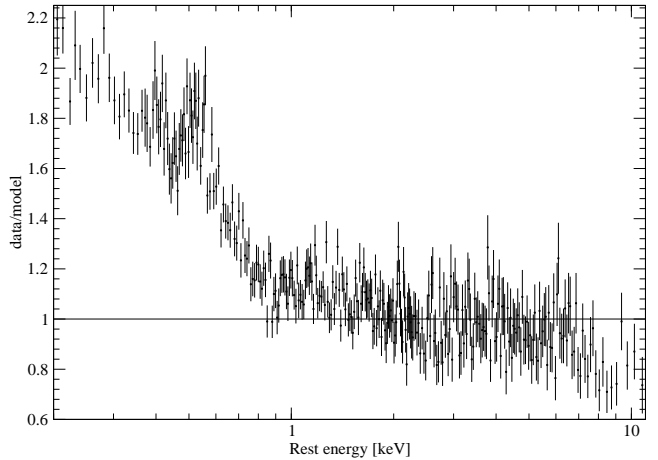}}
\caption{Ratio of the 2000 {\it XMM-Newton} EPIC-pn data to the scaled
2002 EPIC-pn model. The model was scaled by multiplying the power law
and the ionization parameter of the high-ionization WA by a factor
of 1.9.  The additional soft excess in the 2000 spectrum is evident.
\label{2002scale2000} }
\vspace{-0.3cm}
\end{figure}

As argued above, a full model for the X-ray spectrum of \mr\
obtained by {\it XMM-Newton} in 2002 requires a a power law continuum
attenuated by (neutral) Galactic absorption and at least two absorbers.
In Figure~\ref{2002scale2000} we plot the EPIC-pn 2000 data divided
by the scaled 2002 model. The scaling is done by multiplying the
power law continuum flux and the ionization parameter of the WA by
the same factor of 1.9 (which is the hard flux ratio between the
two observations). The plot shows a large excess at low energies
indicating the presence of an additional continuum component. We
denote this continuum the ``soft excess''. We have re-visited the
2002 observation in attempt to look for this component but the data
do not require its presence in this observation.

Next we attempted to determine the shape of the high energy
continuum during the 2000 observation. For this we first fitted
the 3--11 keV band (excluding the 4.5--7.5 keV rest-frame band)
with a Galactic absorbed power law. We find $\Gamma=1.656\pm0.055$
and normalization of $(7.60^{+0.2}_{-0.34})\times10^{-3}$
photons\,cm$^{-2}$\,s$^{-1}$\,keV$^{-1}$ with $\chi^{2}_{\nu}=0.99$.
Fixing the hard continuum at these values, we re-fitted the data
adding this time the ionized and the neutral absorbers, and the
ionized emission constrained to the ionized absorber. We have
also included the additional soft excess component assuming
it can be fitted by a second power law. The fit results
with a $\chi^{2}_{\nu}=1.2$ and the following parameters:
A soft excess component with $\Gamma=2.92^{+0.16}_{-0.16}$
and normalization of $(2.20^{+0.22}_{-0.20})\times10^{-3}$
photons\,cm$^{-2}$\,s$^{-1}$\,keV$^{-1}$, an ionized absorber
with log($U_{\rm OX})= -1.72^{+0.06}_{-0.04}$ and $N_{\rm
H}=10^{21.57\pm0.05}$, and a neutral absorber with a column density
of $10^{20.3^{+0.2}_{-0.6}}$ \cmii .

\subsection{Analysis of the 2000 RGS Spectrum}
\label{2000-rgs}

Although the source flux during the 2000 {\it XMM-Newton} observation
was a factor of $\sim 2$ larger than the flux during the 2002
observation, the integration time is much shorter (by a factor of 5).
Thus, we were unable to obtain any useful constraints from the RGS
data obtained in 2000. More specifically, we could not identify any
emission or absorption lines in this spectrum (if any such lines are
present in this spectrum and have the same EWs as the lines in the
2002 observation, they are consistent with the noise level).

\begin{deluxetable}{cccc}
\tablecolumns{4}
\tablewidth{200pt}
\tablecaption{Hard continuum X-ray power laws
\label{slopes}}
\tablehead{
\colhead{Mission name} &
\colhead{Date} &
\colhead{$\Gamma$} &
\colhead{Normalization\tablenotemark{a}}
 }
\startdata
{\it ASCA}      & 1993 Nov 11 & $1.68^{+0.13}_{-0.12}$ & $1.25^{+0.25}_{-0.20}$ \\
{\it ASCA}      & 1993 Nov 16 & $1.57^{+0.09}_{-0.09}$ & $0.91^{+0.12}_{-0.11}$ \\
{\it ASCA}      & 1993 Dec 12 & $1.66^{+0.08}_{-0.08}$ & $1.15^{+0.13}_{-0.12}$ \\
{\it ASCA}      & 1993 Dec 14 & $1.60^{+0.09}_{-0.09}$ & $0.95^{+0.13}_{-0.11}$ \\
{\it ASCA}      & 1993 Dec 19 & $1.56^{+0.10}_{-0.10}$ & $0.81^{+0.13}_{-0.11}$ \\
{\it ASCA}      & 1993 Dec 24 & $1.57^{+0.11}_{-0.10}$ & $0.69^{+0.11}_{-0.09}$ \\
{\it ASCA}      & 1996 May 26 & $1.42^{+0.12}_{-0.08}$ & $0.43^{+0.05}_{-0.05}$ \\
{\it ASCA}      & 1996 Jun 18 & $1.45^{+0.10}_{-0.07}$ & $0.56^{+0.06}_{-0.05}$ \\
{\it ASCA}      & 1996 Nov 27 & $1.39^{+0.15}_{-0.07}$ & $0.43^{+0.05}_{-0.05}$ \\
{\it ASCA}      & 1996 Dec 09 & $1.35^{+0.14}_{-0.07}$ & $0.43^{+0.05}_{-0.05}$ \\
{\it BeppoSAX}  & 1998 Jun 14 & $1.62^{+0.04}_{-0.04}$ & $0.84^{+0.05}_{-0.05}$ \\
{\it BeppoSAX}  & 1998 Nov 12 & $1.65^{+0.04}_{-0.05}$ & $0.92^{+0.06}_{-0.05}$ \\
{\it XMM-Newton}& 2000 May 29 & $1.66^{+0.06}_{-0.06}$ & $0.76^{+0.20}_{-0.34}$ \\
{\it XMM-Newton}& 2002 May 18 & $1.54^{+0.02}_{-0.02}$ & $\phn0.40^{+0.10}_{-0.14}$ 
\enddata
\tablecomments{Fits to the hard X-ray continuum (3--10 keV, excluding
the rest frame 5.0--7.5 keV band) using a simple power law.}
\tablenotetext{a}{Power law normalization at 1 keV in units of $10^{-2}$ ph\,cm$^{-2}$\,s$^{-1}$\,keV$^{-1}$.}
\end{deluxetable}

\subsection{Historical Variations of the X-ray Spectrum of \mr}
\label{history}

The 2000 and 2002 EPIC-pn observations of \mr\ show that a full model
for the X-ray spectrum must include a high energy power law continuum,
a soft excess power law component and two absorbers. This combination
was used to fit also the earlier {\it BeppoSAX} and {\it ASCA}
spectra of the source and the results are discussed in this section.
We note that the low resolution {\it BeppoSAX} and {\it ASCA} data are
not sensitive to the inclusion of the emission component in our model.
These data are also not sensitive to the differences between the two
and the three absorption components discussed in \S~\ref{rgsfit}.
Thus in the fit below we include only one highly ionized WA component
which represent an average of two such components.

Our initial assumption is that the absorbers' column densities are
constant and that the only changes are in the value of the ionization
parameter which is proportional to the source luminosity.  We fixed
the power law slopes of the soft excess and the hard continuum
to be $\Gamma=2.9$ and $\Gamma=1.6$, respectively, and we only
allow changes in their relative normalization.  The $\Gamma=1.6$
value for the hard X-ray slope is consistent with all previous X-ray
observations of \mr\ and is in accord with earlier findings of Reeves
\& Turner (2000), Orr et al. (2001), and Morales \& Fabian (2002).
In Table~\ref{slopes} we show power-law fits for the 3--10 keV band
(excluding the rest frame 5.0--7.5 keV band) for all data sets in
order to determine hard X-ray slope in each observation.  As seen
from the table, a $\Gamma=1.6$ continuum is in good agreement with all
data sets.  We also fixed the Galactic absorption to the value found
earlier. In the following fits we excluded the rest frame 5.0--7.5
keV band to avoid complications due to the Fe\,K$\alpha$ line. The
bound-free opacity of the low-ionization absorber is very similar to
a totally neutral absorber, and cannot be distinguished from such an
absorber in the data collected with low spectral resolution. Indeed no
useful constraints regarding changes in the ionization state of this
component can be obtained. Thus, in this section we simply approximate
the effects of the low-ionization absorber by a neutral absorber
with a  column density to $10^{20.3}$\,\cmii\ (as found above).
Constraints can, however, be obtained by the study of variations in
the high-ionization absorber and underlying continuum, and these are
considered in the remainder of this section.

\begin{figure}
\centerline{\includegraphics[width=8.5cm]{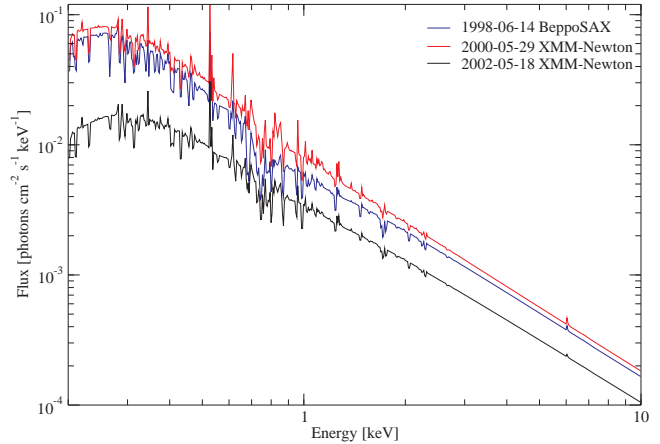}}
\caption{Adopted models for the {\it XMM-Newton} and {\it BeppoSAX}
observations of \mr . Models show the soft excess, the WA, the emission
line gas, and the hard power law, but do not include the iron-K$\alpha$
line and the Galactic absorption for clarity. Note the disappearance
of the soft excess in the 2002 {\it XMM-Newton} observation and the
difference in the absorption around 1 keV.
\label{softxmodels} }
\end{figure}

We first fit the {\it BeppoSAX} data, and re-fitted the {\it
XMM-Newton} data, with the same model and the fixed parameters
as described above. The results of these fits are tabulated
in the upper part of Table~\ref{xrayfit} and are plotted in
Figure~\ref{softxmodels}. The fits to the {\it XMM-Newton}
observations indicate that the flux decrease from 2000 to 2002 was
associated with the disappearance of the soft excess component. On
the other hand, there is no change, within the data and model
uncertainties, in the column density ($N_{\rm H}=10^{21.5-21.8}$)
and the ionization parameter of the high-ionization WA. Thus, there
seems to be no connection between the ionization parameter and the
X-ray luminosity of the source during this two year period. For the
two {\it BeppoSAX} observations, we found that the required WA column
density is $10^{21.8{\pm0.1}}$ \cmii . This is consistent with the
column density reported by Orr et al. (2001) and gives a better fit
($\chi^{2}_{\nu}$ is lower by 0.3) than the column of $10^{21.5}$ \cmii
.  This is significant at an F-test probability of $3\times10^{-7}$.
Fitting the 0.1--11 keV {\it BeppoSAX} observations we find that
they also require the presence of a soft excess component. The two
observations are entirely consistent with each other (as reported
also by Orr et al. 2001) despite the 5 months separation. The high
energy flux level of the {\it BeppoSAX} and the 2000 {\it XMM-Newton}
observations are similar but the soft excess component contribution to
the {\it BeppoSAX} observation is lower by a factor of $\sim 2$ (see
Figure~\ref{softxmodels}).  The WA properties of the {\it BeppoSAX}
observations are not consistent with those of the {\it XMM-Newton}
observations, indicating again that the absorber in \mr\ is changing
on timescales of years. In general, we find a weakening in the soft
excess as the source luminosity decreases, however, the dependence
between the two is not simple.

\begin{figure}
\centerline{\includegraphics[width=8.5cm]{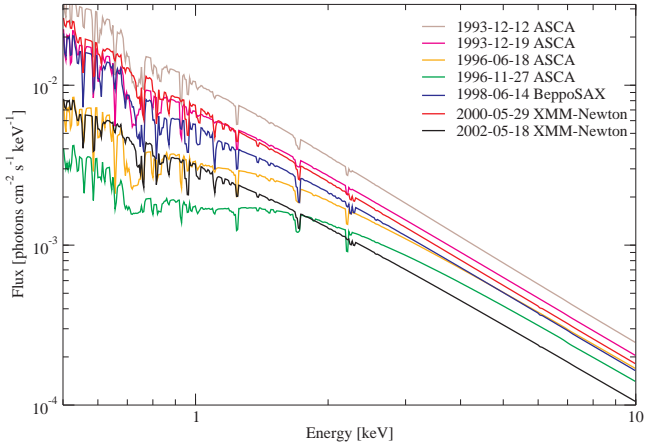}}
\caption{Adopted models for selected X-ray observations of \mr. Models
are showing the WA and power law and do not include the emission line
gas or the iron-K$\alpha$ line for clarity. The models parameters
are detailed in the lower part of Table~\ref{xrayfit}. The {\it ASCA}
1996 models are corrected for the SIS0 degradation.
\label{xmodels} }
\end{figure}

\begin{figure}
\centerline{\includegraphics[width=8.5cm]{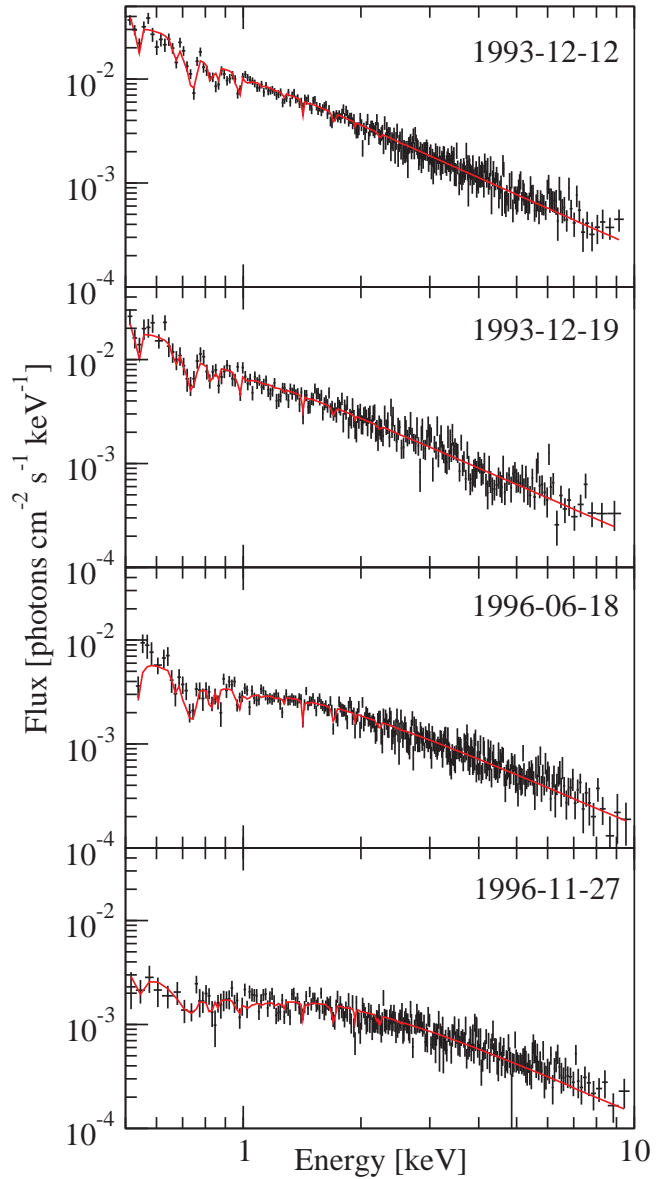}}
\caption{Selected {\it ASCA} SIS0 data sets shown together with the models
described in Table~\ref{xrayfit}.
\label{ascadata} }
\end{figure}

We now consider the {\it ASCA} data, which are limited to the
0.5--10.0 keV band. Therefore we re-fitted the {\it BeppoSAX} and
{\it XMM-Newton} data over the same energy range thus enabling a
more meaningful comparison between the various observations. The
lower part of Table~\ref{xrayfit} summarizes the fit results and
Figure~\ref{xmodels} shows selected models. In Figure~\ref{ascadata}
we show selected {\it ASCA} SIS0 data sets.

We were able to fit the first 6 {\it ASCA} observations (taken during
6 weeks on 1993 Nov--Dec) with the same column density high-ionization
WA ($10^{21.5}$ \cmii ) changing only the ionization parameter (see
Table~\ref{xrayfit}). However, the 1996 Nov-Dec observation requires
a larger column density of $10^{22.1\pm0.05}$ \cmii\ which gives
$\chi^{2}_{\nu} = 1.01$.  When fixing the column density to $10^{21.5}$
\cmii\ the best fit gives $\chi^{2}_{\nu} = 1.76$, i.e., the change
in column density is highly significant (an F-test probability of
practically 0). The best fit to the May-June 1996 observations also
requires a larger column density absorber ($10^{21.84\pm 0.06}$
\cmii ) which gives $\chi^{2}_{\nu} = 1.03$ (when fixing the column
density to $10^{21.5}$ or $10^{22.1}$ \cmii\ the $\chi^{2}_{\nu}$
is increased by $\sim 0.15$ over the 1020 degrees of freedom). This
suggests that the properties of the WA have changed between 1993 and
the end of 1996. The four observations of 1996 can be divided into
two groups: two observations dating 1996 May--Jun and two in 1996
Nov--Dec. Within each group the spectra are indistinguishable. However,
we could not find a consistent model for the two epochs together.

The differences between 1993 and 1996, and the differences during
1996, all indicate that there are real changes in the absorbing
column on timescales of several months to several years and there
are no differences on timescales of two months or less. The {\it
ASCA} data are limited to energies above 0.5 keV and thus poorly
constrain the soft excess component. This also introduces some
uncertainty concerning the WA properties since we cannot unambiguously
determine the contribution of the soft excess component. Like the
{\it XMM-Newton} and the {\it BeppoSAX} results, the {\it ASCA} fits
also indicate a general trend where the soft excess is stronger when
the hard flux is higher.

\section{Spectral Analysis of the UV Data}

\subsection{The High Resolution {\it FUSE} Spectrum}

The FUSE spectrum of \mr\ (Figure~\ref{fusespec}) shows broad
emission lines of \ion{O}{6}$\,\lambda\lambda$1032,1038 and
\ion{C}{3}\,{$\lambda$977}. All these lines show significant
blueshifted absorption. We also detect blueshifted absorption
from at least 10 lines of the \ion{H}{1} Lyman series starting
with Ly$\beta$ and going up to the Lyman edge, where the lines are
blended together. The Ly$\alpha$ absorption is outside the {\it FUSE}
wavelength range and was observed, independently, by {\it HST}.

\begin{figure}
\centerline{\includegraphics[width=8.5cm]{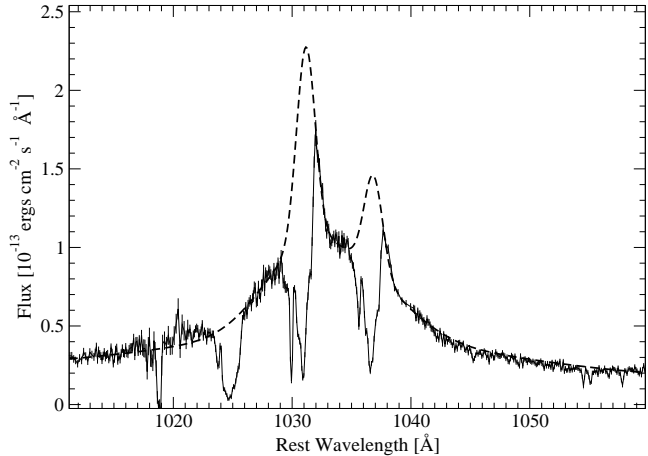}}
\caption{Fit to the \ion{O}{6} emission lines using 3 Gaussians for
each line (dashed line) plotted over the {\it FUSE} spectrum (solid
line) which is binned to $\sim 0.1$ \AA). The fit parameters are
detailed in Table~\ref{o6fittab}.
\label{o6fit} }
\end{figure}

\begin{deluxetable}{cccc}
\tablecolumns{4}
\tablewidth{200pt}
\tablecaption{\ion{O}{6} fit parameters
\label{o6fittab}}
\tablehead{
\colhead{Gaussian} &
\colhead{Normalization\tablenotemark{a}} &
\colhead{$\sigma$\tablenotemark{b} \ [\AA]} &
\colhead{FWHM\tablenotemark{b} \ [\kms]}
 }
\startdata
I   & 1.240  &  \phn 0.782  & \phn 530  \\
II  & 0.518  &  \phn 3.667  & 2500 \\
III & 0.160  &  13.096 & \phn\phn8950 
\enddata
\tablecomments{All Gaussian centers were set to the same velocity. In
the rest frame spectrum the source they correspond to 1031.15\,\AA\
and 1036.84\,\AA , i.e., the fitted \ion{O}{6} emission lines are
blueshifted by $-$240\,\kms\ relative to the optical lines.}
\tablenotetext{a}{The \ion{O}{6}$\lambda$1032 line normalization
in units of $\times10^{-13}$ \ergcmsA . The \ion{O}{6}$\lambda$1038
line normalization was set to half the \ion{O}{6}$\lambda$1032
line normalization.}
\tablenotetext{b}{$\sigma$ and the FWHM are the same for the two lines
of the doublet.}
\end{deluxetable}

\begin{figure}
\centerline{\includegraphics[width=8.5cm]{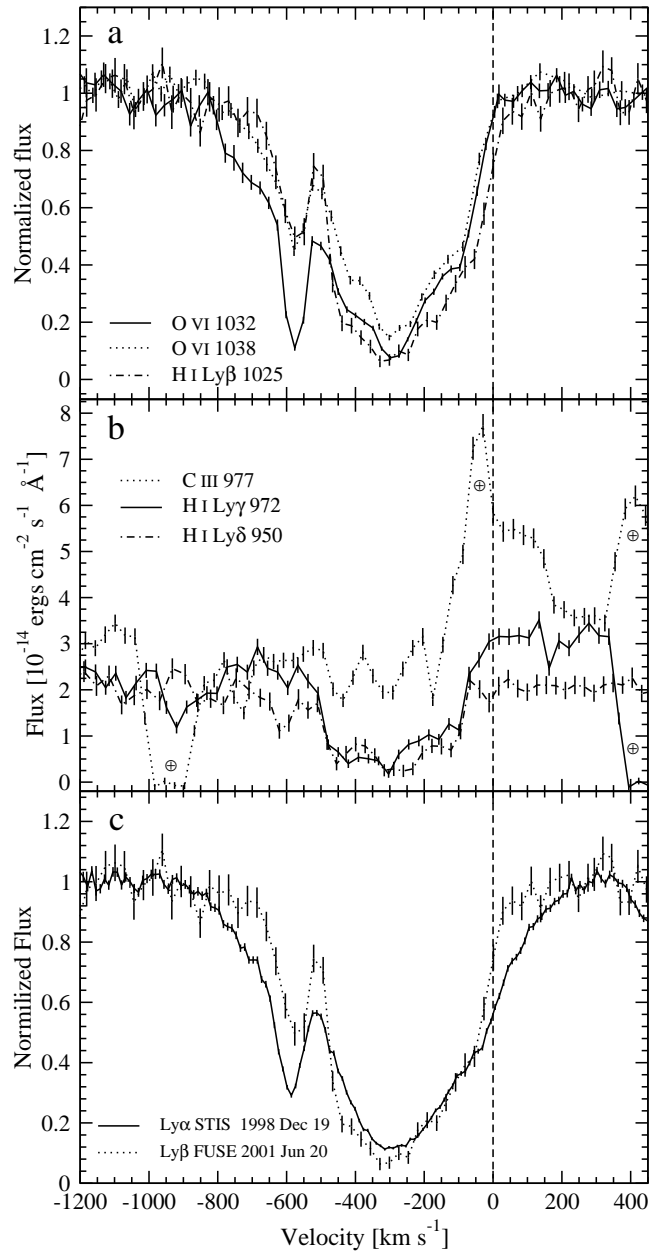}}
\caption{(a) Velocity spectra for the two \ion{O}{6} lines 
and for Ly$\beta$. 
(b) Velocity spectra for \ion{C}{3}, Ly$\gamma$, and Ly$\delta$.
Airglow and Galactic absorption and emission lines are marked
with~$\oplus$.
(c) Velocity spectra for Ly$\alpha$ observed with STIS on 1998 December
19 (solid line) and for Ly$\beta$ observed with {\it FUSE} on 2001 June
20 (dotted line). 
\label{velspec} }
\end{figure}

In order to study the intrinsic absorption spectrum we first fitted
the \ion{O}{6} doublet emission lines. Each of the two emission
lines ($\lambda\lambda$1032,1038) was fitted with three kinematic
components represented by Gaussians with the same kinematic width,
and had their flux ratio fixed at the ratio of oscillator strengths
(2:1).  The results are shown in Figure~\ref{o6fit} and are listed
in Table~\ref{o6fittab}. The observed spectrum was then divided by
this emission model and the resulting normalized spectrum was used to
obtain the absorption velocity spectra shown in Figure~\ref{velspec}a.
We note that the continuum shown in Figure~\ref{o6fit} does not match
very well the continuum on the blue side of the Ly$\beta$ line around
1021~\AA . To produce the velocity spectrum of Ly$\beta$ (shown in
Figure~\ref{velspec}a), we fitted the continuum on both sides of the
line with a spline curve and divided the observed spectrum by this
continuum. Figure~\ref{velspec}b shows the \ion{C}{3} absorption
and two more Lyman absorption lines. The \ion{C}{3} line region
contains many Galactic features which do not allow a proper continuum
fit. Therefore, this region was not normalized. Nevertheless, there
is a clear \ion{C}{3} absorption which matches in its velocity range
the blueshifted absorption seen in Figure~\ref{velspec}a.

Figures~\ref{velspec}a~and~\ref{velspec}b suggest that the intrinsic
absorption in \mr\ is arising in at least 4 absorption systems: one
at $-$580 \kms\ and at least 3 others which are blended together and
form a wide trough covering the velocity range 0 to $-$500 \kms .
The 3 centroid velocities in the trough are at about $-150, -300$,
and $-430$ \kms .

\subsection{The {\it HST} UV Spectrum}
\label{hst}

UV spectra of \mr\ were taken by {\it HST} at three epochs (Monier et
al. 2001 and references therein): 1996 August 2 with the FOS, 1998
December 19 and 2000 November 5 with STIS. To produce the velocity
spectrum of the Ly$\alpha$ absorption line, we fitted the 1998 STIS
spectrum with a spline curve and divided the observed spectrum by this
continuum. The Ly$\alpha$ absorption line shows a similar profile to
the Ly$\beta$ profile observed with {\it FUSE} 2.5 years later (see
Figure~\ref{velspec}c). The similarity is mainly in the center of the
lines while the later spectrum suggesting the absorption got narrower
(though this could be an artifact of the continuum fitting). The
similarity in the lines centers indicates they are saturated.  However,
the lines are not completely black indicating the absorber does not
completely cover the continuum source. The line profile suggests a
covering factor of $\sim$90\% for the absorbing material. This value
is consistent with the covering factor fitted to the X-ray data.
The exception is the narrow trough at $\sim-580$ which has a different
depth in the two lines.  One possibility is that, in this system,
the \ion{H}{1} lines are not saturated. The alternative explanation
is that the system is saturated but it lies under the blue wing of
the main broad absorption trough and this broader absorption is not
saturated, resulting in a different depth to the blue wing.

Ganguly, Charlton, \& Eracleous (2001) suggested that the \ion{C}{4}
doublet absorption line that was detected in the FOS observation,
in 1996 with EW of 1.09$\pm$0.09 \AA , was not detected in the STIS
spectrum taken 4 years later (down to a 3$\sigma$ limit corresponding
to EW=0.19 \AA ). The \ion{C}{4} absorption system detected in the
FOS spectrum has a width of $\sim 400$ \kms\ (Monier et al. 2001),
consistent with the large trough seen in the {\it FUSE} spectrum. The
resolutions of the two spectra are too poor (230 and 600 \kms\ for
the FOS and STIS, respectively) to detect the narrow absorption system
at $-580$~\kms\ (which has a width of order 100 \kms).

\section{Discussion}

\subsection{The Highly Ionized Gas in \mr}
\label{historydiss}

The new {\it XMM-Newton} data presented in this paper clearly show
the presence of ionized gas seen in both emission and 
absorption in \mr .
The spectral analysis is severely limited by the poor S/N of
the grating observations but several interesting results clearly
emerge.  In the grating data we identified X-ray emission lines from
\ion{N}{6}, \ion{O}{7}, \ion{O}{8}, \ion{Ne}{9}, and \ion{Ne}{10}. We
also identify, with high certainty, absorption lines from the
low ionization ions of \ion{O}{3}, \ion{O}{4}, and \ion{O}{5},
as well as the signature of absorption edges due to \ion{O}{7}
and \ion{O}{8}. Many other absorption lines are probably detected
(Table~\ref{linetable}) but their reality and intensities are highly uncertain,
because of the limited S/N.

For the 2002 RGS data we suggest one of two possible models. The first
model consists of two absorbers: a highly ionized absorber with a
column density of $10^{21.5-21.8}$ cm$^{-2}$  and log($U_{\rm OX})=-1.68$,
and a low ionization absorber with a column density of $10^{20.3}$
cm$^{-2}$ and log($U_{\rm OX})=-4.00$.
The second possibility is a three-component model where we split the
highly ionized absorber from the above model into two components: one
with log($U_{\rm OX})=-1.4$ and the other with log($U_{\rm OX})=-2.6$.
The ionized
(line-of-sight) absorber and the ionized emitter in both cases are
consistent with being the same gas with a global covering factor of
0.4. The highly ionized absorption lines are probably narrower than
200 \kms , a limit which is imposed by the equivalent width of the
strongest predicted lines, given the column density, the ionization
parameter, and the S/N of the observations.

More interesting conclusions are derived from analyzing the historical
light curve of the source combining 8.5 years of observations by {\it
XMM-Newton}, {\it ASCA} and {\it BeppoSAX}. The main findings are:
\begin{enumerate}
\item
All X-ray observations are consistent with a two-component continuum:
a high energy power law of slope $\Gamma = 1.6$ and a low energy soft
excess component with $\Gamma = 2.9$. Both components are absorbed
by the WA and by the intrinsic neutral gas.
\item 
The WA observed during the 6 weeks of {\it ASCA} observations in 1993
is consistent with being a single absorber with a column density
of $10^{21.5}$ \cmii . Less conclusive results are obtained for
the short timescales behavior due to the poor S/N.  The data are
consistent with a scenario in which the decrease in flux caused a
corresponding decrease in the ionization parameter. Such a behavior
has been suggested in the past for several other sources (e.g.,
MCG$-$6$-$30$-15$ -- the {\it ASCA} observation of Otani et al. 1996;
NGC\,3516 -- the {\it Chandra} observation of Netzer et al. 2002).
This interpretation is not unique and the data also supports a more
complex case where the source luminosity is not simply correlated with
the ionization parameter (e.g., MCG$-$6$-$30$-15$ -- Orr et al. 1997;
NGC\,3783 -- Behar et al. 2003; Netzer et al. 2003)
\item 
On timescales of years, the WA properties are different and our model
requires that the absorbing gas properties are changing in time. For
example, the {\it ASCA} 1996 observations clearly indicate a larger
column density ($10^{21.8}$ \cmii\ vs. $10^{21.5}$ \cmii ) and a
smaller ionization parameter (log$(U_{\rm OX}) \sim -2.3$ vs. $\sim
-1.5$) absorber compared with the one observed in 1993. This could
indicate new material entering our line-of-sight, between 1993
and 1996, adding to or replacing the earlier gas. The two groups
of observations taken in 1996, that are separated by 5 months, are
also not consistent with the notion of having the same WA. Comparing
these two periods we find that the luminosity is about the same while
the ionization parameter dropped by about a factor of 3. We suggest,
again, a physical motion of the gas which resulted in a higher column
density of material at about the same distance. Thus, a real change
in the absorber properties can take place over time scales of only
a few months.

The comparison of the 1993 {\it ASCA} observations and the 2002
{\it XMM-Newton} observation suggest a different change. In 2002,
the AGN flux is smaller by a factor of 2 compared with 1993,
yet the column density and the derived ionization parameter are
about the same as in 1993. Similarly, a comparison between the two
{\it XMM-Newton} observations indicates that the source luminosity
decreased significantly from 2000 to 2002 yet the derived WA properties
remained about the same. This might mean that the absorbing material
properties have changed between the two epochs (the SED in both is
very similar but the luminosity decrease between 2000 and 2002 was
not accompanied by a corresponding decrease in ionization parameter).
An alternative explanation is that the absorbing material is very
far from the central source and of low enough density such that it
did not respond to the continuum luminosity variations.
\item
The soft excess continuum luminosity is positively correlate with
the hard continuum luminosity.
\end{enumerate}

The overall picture which emerges from this study is of a changing
absorber made of material that enters and disappear from the
line-of-sight on timescales of several months. On shorter timescales,
of several weeks, the models are consistent with a picture in which
the absorbing material responds instantly to the continuum luminosity
variations. Due to the data quality and the model complexity (two
power laws and several absorbers) we cannot unambiguously determine
those properties.

\subsection{The ``Neutral'' Absorber}

The 2002 {\it XMM-Newton} observation show the presence of a
low-ionization/neutral absorber intrinsic to \mr. We derived the
column density of this gas ($10^{20.3}$ \cmii) from the shape of the
soft X-ray continuum below 0.6 keV. The RGS data show evidence for
\ion{O}{3} and \ion{O}{4} absorption lines which allow us to constrain
its level of ionization (log$(U_{\rm OX})=-4.0$).

Macchetto et al. (1990) find evidence for circumnuclear gas on
distances between 3 and 6 kpc and of gaseous filaments farther out at
distances of 30--50 kpc. Their lower limits on the [\ion{O}{3}] density
(e.g., their table 4) implies a column densities of $\sim10^{20}$
\cmii . This is entirely consistent with the properties of the low
ionization absorber found in our analysis. Thus, it is possible that
the same gas responsible for the [\ion{O}{3}] emission is detected in
absorption via \ion{O}{3} and \ion{O}{4} X-ray absorption lines. This
means that parts of the emission line nebula observed in \mr\ lie in
our line of sight to the central source.

\subsection{Ultraviolet Emission and Absorption}
\label{hiresfuse}

This paper presents the first FUV spectrum of \mr . We detect emission
from \ion{O}{6}, \ion{N}{3}, and \ion{C}{3}. We also detect at least
4 absorption systems in \ion{O}{6}, \ion{C}{3}, and \ion{H}{1},
three of which are blended together. The three blended systems are
best seen in \ion{C}{3} (Figure~\ref{velspec}b) and are definitely
suggested in the other absorption lines (the main absorption system
at $\sim -300$\,\kms\ looks like 3 systems blended together, see
Figure~\ref{velspec}a).

The \ion{H}{1} Lyman absorption lines are seen all the way to the
Lyman edge (Figure~\ref{fusespec}). The lower series lines have
similar EWs which suggests saturation. However, the lines are not
completely black which means incomplete line of sight coverage.

Figure~\ref{velspec}a demonstrate that the \ion{O}{6} absorption
system with the largest blueshift might not be saturated since the
depth of the 1038\,\AA\ line is about half the depth of 1032\,\AA ,
as expected from their oscillator strength ratios. On the other hand,
the three blended absorption systems have the same depth in both lines
of the \ion{O}{6} doublet, indicating that these systems are saturated.
The similarity of the \ion{H}{1} and the \ion{O}{6} absorption profiles
suggests that all blended lines of the two ions are saturated and
have the same covering fraction. Our spectral analysis indicates a
covering factor between $\sim$60\% and $\sim$90\% for these systems.

The total depth in all the FUV and UV absorption lines is larger than
the underlying continuum, thus the broad emission lines are absorbed
by the UV absorber. This indicates that the UV absorber lies outside
of the BLR. Using \mr\ continuum flux at 5100~\AA\ ($(37\pm 3)\times
10^{-16}$ \ergcmsA ; Bergeron et al. 1983) and the relation between the
BLR size and the object's luminosity (equation 6 of Kaspi et al. 2000)
we find the BLR size to be $(1.2\pm 0.2)\times 10^{17}$ cm. We take
this to be a lower limit on the distance of the UV absorber from the
central black hole.

\subsection{The Ultraviolet --- X-ray Connection}

Several studies suggested a link between the UV and X-ray absorber
in AGNs (e.g., Mathur et al 1994; Mathur, Elvis, \& Wilkes 1995;
Shields \& Hamann 1997; Crenshaw et al. 1999). The data presented in
this paper enables us to study this connection in \mr .

As explained, the resolution and S/N of the RGS spectrum does not allow
the exact measurement of the X-ray absorption systems. We can only
confirm that the velocity shifts and the FWHM of the UV absorption
systems are consistent with the ones observed in the X-ray absorber.
We found that the FUV absorption lines (in \ion{H}{1}, \ion{C}{3},
and \ion{O}{6}) are blends of at least 4 absorption systems, all
blueshifted with respect to the emission line by 0 to -600 \kms
. The {\it HST} spectra (\S~\ref{hst}) show a Ly$\alpha$ absorption
line which is consistent with the {\it FUSE} \ion{H}{1} absorption
lines. It also shows a \ion{C}{4} absorption which has a width of
400 \kms\ and about the same blueshift range as seen in the FUV.
The resolution of the RGS spectrum at 22 \AA\ is $\sim$550 \kms\
hence any absorption lines which are similar in width to the UV
absorption lines are predicted to be one pixel wide. In the X-ray
spectrum we identify \ion{O}{3}, \ion{O}{4}, and \ion{O}{5} which are
consistent with the widths and blueshifts of the UV absorption lines,
as well as hints for absorption from \ion{O}{6}. We identify few
absorption lines from highly ionized species with lines at wavelength
shorter than 19~\AA\ (see Table~\ref{linetable}). Given the poor
S/N, this is consistent with the absorption seen in the UV spectra.
The consistency between the UV and X-ray absorption suggests that
they could arise in the same gas.

We have set a lower limit of $(1.2\pm 0.2)\times 10^{17}$ cm on the
distance of the UV absorber from the source using the BLR distance.
If the UV and X-ray absorptions are the same, this is also a lower
limit on the distance of the X-ray absorber.

\acknowledgments

We thank E. Behar for helpful discussions.  We also thank the anonymous
referee for constructive comments.  We acknowledge a financial support
by the Israel Science Foundation grant no. 545/00. H. N. thanks the
Columbia University astrophysics group for their hospitality and
support during part of this investigation. This research has made
use of the NASA/IPAC Extragalactic Database (NED) which is operated
by the jet Propulsion Laboratory, California Institute of Technology,
under contract with the National Aeronautics and Space Administration;
the Abstract Service of NASA's Astrophysics Data System; data obtained
through the HEASARC on-line service, provided by NASA/GSFC; and data
from the TARTARUS database, which is supported by Jane Turner and
Kirpal Nandra under NASA grants NAG5-7385 and NAG5-7067.

\end{document}